# Hybrid deep neural network based prediction method for unsteady flows with moving boundaries

Renkun Han[1], Zhong Zhang[2], Yixing Wang[1], Ziyang Liu[1], Yang Zhang[1] and Gang Chen[1, a)]

[1]*State Key Laboratory for Strength and Vibration of Mechanical Structures, shannxi Key Laboratory for Environment and Control of Flight Vehicle, School of Aerospace Engineering, Xi'an Jiaotong University, Xi'an 710049, China*

[2]*Science and Technology on Reliability and Environment Engineering Laboratory, Beijing Institute of Structure and Environment Engineering, Beijing 100076, China*

A novel hybrid deep neural network architecture is designed to capture the spatial-temporal features of unsteady flows around moving boundaries directly from high-dimensional unsteady flow fields data. The hybrid deep neural network is constituted by the convolutional neural network (CNN), improved convolutional Long-Short Term Memory neural network (ConvLSTM) and deconvolutional neural network (DeCNN). Flow fields at future time step can be predicted through flow fields by previous time steps and boundary positions at those steps by the novel hybrid deep neural network. Unsteady wake flows around a forced oscillation cylinder with various amplitudes are calculated to establish the datasets as training samples for training the hybrid deep neural networks. The trained hybrid deep neural networks are then tested by predicting the unsteady flow fields around a forced oscillation cylinder with new amplitude. The effect of neural network structure parameters on prediction accuracy was analyzed. The hybrid deep neural network, constituted by the best parameter combination, is used to predict the flow fields in the future time. The predicted flow fields are in good agreement with those calculated directly by computational fluid dynamic solver, which means that this kind of deep neural network can capture accurate spatial-temporal information from the spatial-temporal series of unsteady flows around moving boundaries. The result shows the potential capability of this kind novel hybrid deep neural network in flow control for vibrating cylinder, where the fast calculation of high-dimensional nonlinear unsteady flow around moving boundaries is needed.

## 1. Introduction

Any motion, forced or free, of boundaries in steady approach flow clearly affects the flow field in the wake of the boundaries (Atluri *et al.*, 2009). The flow around oscillating boundaries is an important engineering problem both from the academic and practical points of view (Sarpkaya *et al.*, 2004). Examples are chimney stacks, transmission lines, cables of suspended bridges, offshore structures and risers, which are exposed to wind or ocean currents. The practical significance of this type of flow has led to a large number of fundamental studies. As one important part of these studies, high-fidelity modeling of unsteady flows around oscillating boundaries is one of the major challenges in flow control or other applications. The high-fidelity computational fluid dynamics (CFD) techniques have made significant inroads into this problem (Facchinetti *et al.*, 2004). Depending on billions of degrees of freedom and dynamic mesh, the computational cost is unbearably high, and it is practically difficult to purely rely on the time domain CFD simulation for fast flow control. Therefore, many research efforts have been devoted to data-driven low-dimensional models, which can capture the main dynamic characteristics of unsteady dynamic systems with good efficiency and enough accuracy (Fang *et al.*, 2013).

Reduced order modeling (ROM), such as proper orthogonal decomposition (POD) (Dowell, 1997) and dynamic mode decomposition (DMD) (Schmid, 2010), offers the potential to simulate physical and dynamic systems with substantially increased computational efficiency while maintaining reasonable accuracy. A Lot of researches (Chen *et al.*, 2018; Jovanovie *et al.*, 2014; Hemati *et al.*, 2014) have been carried out to analyze flow fields with low-dimensional representations using these methods. But most of these investigations are linear or weakly nonlinear methods with some strong assumptions, which limits the applications of these methods to more complex unsteady flows. Deep learning technology (LeCun *et al.*, 2015) is a recent advancement in artificial neural networks which is capable of finding more complex and hidden information from the big data. It has advantage of learning the nonlinear system with multiple levels of representation data. Recently, there is a great interest in introducing the deep learning method to fluid mechanics.

The pioneering investigation of using deep learning technology for fluid dynamic is to model some parameter of Reynolds-averaged Navier-Stokes (RANS) equations. (Ling et al., 2016; Wu et al., 2018; Maulik et al., 2019; Xie et al., 2019) These methods increased the accuracy of RANS models by utilizing neural networks to learn Reynolds stress closures. In the work of Wang et al. (2016) and Omata et al. (2019), the deep convolutional autoencoder data-driven nonlinear low-dimensional representation method were used for dimensionality reduction of unsteady flow fields. In those methods, neural networks consist of a convolutional neural network and a deconvolutional neural network. The former is used to capture the spatial features, and the latter is used to reconstruct the high-resolution high-dimensional flow field.

To capture the temporal-spatial features of unsteady flow by deep neural networks, Fukami et al. (2019) and Lee et al. (2019) used CNN to capture spatial features and DeCNN to predict flow fields by the combination of captured features from previous time steps. Pawar et al. (2019) constructed a combination of POD and fully connected neural network. The POD method was used to reduce dimensional size of flow fields and represent flow fields by the combination of POD modals. The fully connected neural network was used to predict the time coefficients of POD modes. Similarly, Rahman et al. (2019), Deng et al. (2019) and Ahmed et al. (2019) used LSTM to predict the time coefficients of POD modals, and Miyanawala et al. (2019) used CNN to predict the time coefficients of POD modals. And combinations of the CNN and ConvLSTM also been proposed for dimensionality reduction and spatial-temporal modeling of the flow dynamics by Mohan et al. (2019), Hasegawa et al. (2019) and Han et al. (2019). All those methods only can predict unsteady flows without moving boundaries. However, moving boundaries make the flow more complex and destructive, which make the flow around moving boundaries an important engineering problem both from the academic and practical points of view.

For a system with moving boundaries, it's very difficult to capture the complex temporal-spatial features of unsteady flow influenced by moving boundaries. Very few studies have used deep neural networks to solve unsteady flow fields with moving boundaries. Raissi et al. (2019) employed deep neural networks that are extended to encode the incompressible Navier-Stokes equations coupled with the structure's dynamic motion equation. This model is able to predict the lift and drag forces on the structure given some limited and scattered information on the velocity field. Srivastava et al. (2019) developed machine learning surrogates based on Recurrent Neural Networks (RNN) for predicting the unsteady aeroelastic response of transonic pitching and plunging wing-fuel tank sloshing system. For the aeroelastic prediction, the inputs are given in form of temporal sequences of airfoil plunge and pitch displacements i.e. $h$ and $\alpha$ and also the network outputs i.e. lift coefficient, $C_L$ and pitching moment coefficient $C_M$ at the previous instant. This model only established the mapping relationship between the boundary position at this moment and forces at the next moment, so that it couldn't show the flow fields in the wake of the body, which means this method couldn't reveal the flow mechanism and couldn't be used in flow control directly.

Based on previous work (Han et al. 2019), a novel hybrid deep neural network is designed to achieve fast and accurate prediction of unsteady flow fields around moving boundaries. This model is able to model the spatial-temporal flow dynamics characteristics at a low computational cost. Different from previous work, we try to change the internal structure of the LSTM cell so that it can simultaneously learn the temporal evolution characteristics of spatial features of flow fields and the influence of boundary position change on the flow field. More specifically, a novel hybrid deep neural network is designed to capture the complex spatial-temporal flow dynamics features directly from the high-dimensional flow fields and boundary position, and then predict the unsteady flow field at future occasions based on the captured features from the flow data at past times and boundary position at future occasions.

The structure of this article is as follows: Sec. II introduces the architecture of the proposed hybrid deep neural network architecture. Then, in Sec. III, the method for constructing flow fields dataset and the neural network training algorithm are explained. Sec IV evaluates the performance of the proposed the hybrid deep neural network. Finally, a summary and conclusion are provided in Sec. V.

## 2. Architecture of the hybrid deep neural network

## 2.1. Architectural design of the hybrid deep neural network

Flow fields at future occasions not only depend on the current state of motion, but also on the time history of motion. Flow fields at future occasions are predicted based on the time history of motion as shown in Eqn. (1), in which $F_i$ is flow fields at time $t_i$ and $G_i$ is grid position at time $t_i$. As explained in section 1, the goal of this work is to reduce the dimension of high dimensional unsteady flow data and to learn its spatial-temporal dynamic characteristics directly from the past time flow fields and boundaries position by deep neural network. Therefore, a hybrid deep neural network architecture composed of CNN layers, ConvLSTM layers and DeCNN layers is designed to capture the spatial-temporal features of unsteady flows, as is shown in Figure 1.

$$F_i = f\left(F_{i-1}, F_{i-2}, ..., F_{i-k}, G_i\right) \tag{1}$$

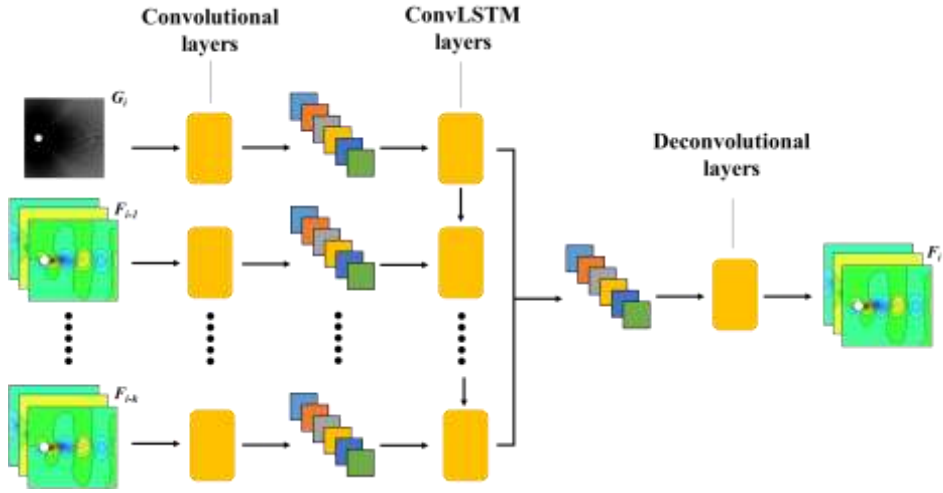

Figure 1. The architecture of the hybrid deep neural network.

The CNN layers are designed to capture complex spatial features directly from the high-dimensional input fields and represent it in low-dimensional form. The convolution operation extracts feature from images, enhancing certain features of the original signal, and reducing noise. Each convolutional layer consists of a set of learnable filters (or kernels), which have a small receptive field, but extend through the full depth of the input volume. By training the network, filters are optimized so that it is able to detect some specific type of feature at some spatial position in the input. After input data flows through CNN layers, several features of each time step flow field are obtained by CNN layers.

The LSTM networks are specialized at capturing the temporal characteristics. The predictions of LSTM networks are conditional on the recent context in the input sequence, not what has just been presented as the current input to the network. For instance, to predict the realization at time $t_i$, the LSTM networks can learn from the data at $t_{i-1}$ and also at $t_{i-k}$, since the outcome of the system depends on its previous realizations, as is shown in Figure 2. To capture deeper information from input data, the number of LSTM layers is increased, as is shown in Figure 3. For the problem of predicting flow fields with moving boundaries, typical LSTM cell is improved to 3D ConvLSTM cell so that it can learn from 3D features obtained from flow fields and boundary position by CNN layers, which will be detailed in the next section. The ConvLSTM layers in this paper are able to capture temporal features between low-dimensional features of flow fields at previous occasions and predict the feature maps of the flow field at future occasions.

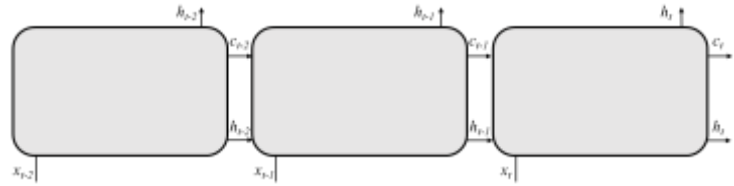

Figure 2. The architecture of a LSTM layout with cell connections.

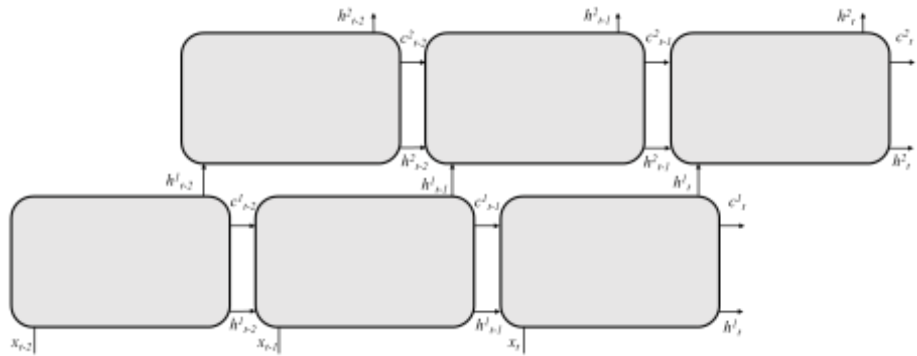

Figure 3. The architecture of a Multi-layer LSTM layout

The essence of DeCNN is convolution, but with an automatic zero padding is added before the convolution. After zero padding, DeCNN layer multiply each element of the input with a filter (kernel) and sum over the resulting feature map, effectively swapping the forward and backward passes of a regular convolutional layer. The effect of using DeCNN layers is to decode low-dimensional abstracted features to a larger dimensional representation. In this paper, the deconvolutional layers copies the architecture of the CNN layers and reverses it. The DeCNN layers represent the predicted low-dimensional feature maps to high-dimensional output flow field, with the same dimension as input fields.

*2.2. Improved convolutional LSTM cell*

A typical LSTM cell contains three gates: the input gate, the output gate and the forget gate. LSTM cell regulates the flow of training information through these gates by selectively adding information (input gate), removing information (forget gate) or letting it through to the next cell (output gate), as is shown in Figure 4. In traditional LSTM cell, the input and hidden states consist of a one-dimensional vector, therefore a two-dimensional input (such as an image or a data field) has to be resized to a single dimension. The "removal" of this dimensionality information fails to capture spatial correlations that may exist in such data, leading to increased prediction errors.

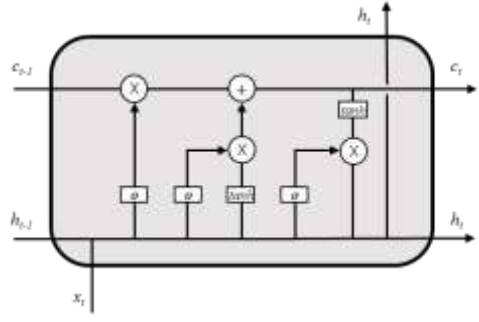

Figure 4. The illustration of the inner structure of ConvLSTM cell.

By setting all gates as 3D tensor and turning point multiplication between vectors into convolutions between tensors, new convolutional LSTM cell is able to capture spatial features from 3D input. This enables us to provide a 3D image input and obtain 3D vectors cell state as outputs from the new ConvLSTM cell. Consider the equations of the improved ConvLSTM cell to compute its gates and states as shown in Eqn. (2). The input gate is represented by $i$, output gate by o and forget gate by $f$. The cell state is represented as $c$ and the cell output is given by $h$, while the cell input is denoted as $x$. The weights for each of the gates are represented as $W$.

$$\begin{aligned}
f_t &= \sigma(W_{xf} * \chi_t + W_{hf} * h_{t-1} + W_{cf} \odot c_{t-1} + b_f) \\
i_t &= \sigma(W_{xi} * \chi_t + W_{hi} * h_{t-1} + W_{ci} \odot c_{t-1} + b_i) \\
c_t &= f_t \odot c_{t-1} + i_t \odot tan(W_{xc} * \chi_t + W_{hc} * h_{t-1} + b_c) \\
o_t &= \sigma(W_{xo} * \chi_t + W_{ho} * h_{t-1} + W_{co} \odot c_t + b_o) \\
h_t &= o_t \odot tan(c_t)
\end{aligned} \quad (2)$$

For the problem of predicting flow fields with moving boundaries, features obtained by CNN layers from boundary position at the time field need to be predicted are set as the cell state $c_{t-1}$ in the first ConvLSTM cell of ConvLSTM layer. Features obtained by CNN layers from flow fields at previous time steps are set as the $x_t$ in the first ConvLSTM cell. Multiple ConvLSTM cells in series to form a ConvLSTM layer, which means that more time steps flow fields are added into this hybrid deep neural network as $x_t$ in following cells.

## 3. Training method

### 3.1. Dataset constructions

Numerical simulations are conducted by solving the nondimensionalized incompressible Navier-Stokes equations as follows:

$$\begin{aligned}
\nabla \cdot u &= 0 \\
\frac{\partial u}{\partial t} + \nabla \cdot uu &= -\nabla p + \frac{1}{Re} \nabla^2 u
\end{aligned} \quad (3)$$

where $u$, $x$, $y$, $t$, $p$ are nondimensionalized velocity, length, time, and pressure, respectively by the incoming velocity $U_0$, the characteristic length $D$, the fluid density $\rho$, and vortex shedding frequency $1/T$. The CFD solver, developed by our research group (Li *et al.*, 2017), had been proven with good accuracy in solving fluid-structure interaction problem. The finite volume method is used in this solver. Lower-Upper Symmetric Gauss-Seidel (LU-SGS) implicit method is employed for time integration. Second-order Van Leer format is employed for spatial discretization. Two-dimensional laminar model is adopted. The dynamic grid method used to update fluid internal mesh in 2D problem are described as following two steps: (1) Based on the displacement of the boundary, the displacement of the grid points on the edge of each block is obtained by one-dimensional infinite interpolation; (2) The displacement of the grid points on the face of each grid block is obtained by the two-dimensional infinite interpolation method.

Since the CNN are developed from the field of computer vision, we consider the goal of predicting flow fields at future occasions based on the time history of flow fields in a similar fashion to approaches considered in deep learning for image-to-image regression tasks. So that the dataset used for training and testing networks should be image like dataset. The flow fields information value at each moment should be distributed over evenly distributed grid points, like pixels. A rectangular area should be chosen as the sampling area. Lattice like sampling points of $N_x \times N_y$ are placed in the space. Then, project the nondimensionalized flow fields variables onto the uniformly distributed grid. The values of points inside the body are 0. Three-dimensional flow field variables ($p^*$, $u^*$ and $v^*$) are extracted at each sampling point. $N_x \times N_y \times 3$ dimensional data are extracted to represent each instantaneous field. The obtained data is arranged in chronological order to obtain a dataset for training and testing the neural network.

*3.2. Training algorithm*

The root mean square error (RMSE) is used to evaluate the model performance, i.e.

$$\text{RMSE}^t = \sqrt{\frac{\sum_{i=1}^{N}\left(\psi_i^t - \psi_{o,i}^t\right)^2}{N}} \tag{4}$$

where $\psi_i^t$ and $\psi_{o,i}^t$ denote the predictions and numerical simulations at the node $i$ and the time level $t$, respectively, and $N$ represents the number of nodes on the full image. Since max error always located at the region near to the moving boundary, the weights factor of the error in the area of twice the diameter are amplified.

Training of the network is carried out with the open-source software library TensorFlow (Abadi *et al.*, 2015). Training the network is equivalent to minimizing the loss function in Eqn. (4) to obtain the optimal all kernel parameters. Adaptive moment estimation (Adam) is employed as the optimization algorithm to train the network (Kingma *et al.*, 2015). In this algorithm, the exponential moving average is used to update the gradient vector and the squared gradient. The whole training procedure is as follows:

(1) Initialize network parameters, including weight *W* and offset *b* for each layer.
(2) Sample a batch in the training dataset, input = {$F_1$, $F_2$, ......, $F_t$, $G_{t+1}$ }, output = {$F_{t+1}$}.

(3) Update iteration step *t+1* flow field $F_{t+1}$.

(4) Compute the RMSE between $F_{t+1}$ and $F_{t+1}$ and get the gradient $g_t$ of the loss function with respect to the parameters.

(5) Update network parameters, $W_{t+1} = W_t - \alpha f(g_t)$, $b_{t+1} = b_t - \alpha f(g_t)$, where *α* is the learning rate.
(6) Repeat Steps 2–5, until the loss function converges.

## 4. Results

*4.1. Training cases*

The near-wake structure of a uniform flow past a circular cylinder undergoing a constant-amplitude transverse forced oscillation is studied most for its wide range of engineering applications. In this article, two cases are used to demonstrate the predictive power of this deep neural network. The training sets are the flow around moving cylinder cases with different amplitudes at Reynolds number equaling to 100 and 200. The calculation domain layout is shown in Figure 5. The distance from the center of the body to the inlet is 20D and to the outlet is 60D. The transverse width of the computational domain is 80D. A Dirichlet boundary condition is imposed on the inlet with the incoming velocity $U_0$; a free out-flow boundary condition is imposed on the outlet; a slip boundary condition is set up for the bottom and top boundaries of the flow; a no-slip boundary condition is set up for the solid body surface. The equation of cylinder motion is defined as:

$$Y = A \times \sin(2 \times \pi \times f \times t) \tag{5}$$

For these two cases, training dataset is the flow around moving cylinder cases with amplitudes A= 0.25D, 0.3D, 0.35D, 0.4D. It is generally expected that a network trained with training samples will have a strong generalization ability, that is, the ability to give a reasonable response to larger range inputs. The case of flow around moving cylinder with amplitude A= 0.425D is used to test the generalization ability of the neural network, whether the hybrid deep neural network could be used to predict the flow field structures beyond the range of training dataset. And the cylindrical vibration frequency *f* is 1.4 *Hz* in all cases.

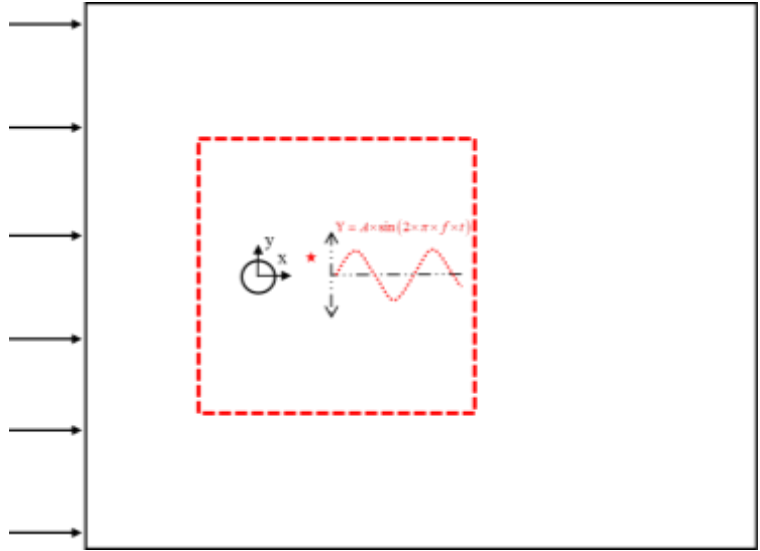

Figure 5. Illustration of calculation domain for oscillating rectangular

To match the deep learning configuration, data used for the network should be preprocessed. The wake flow area in range -1.75D≤x≤8.25D and -5D≤y≤5D is chosen as the velocity field area, area inside the red dotted line showed in Figure 5. Lattice like sampling points of 200 × 200 are placed in the space. Then, project the nondimensionalized flow field variables quantities onto the uniformly distributed grid. The values of the point inside the body are 0. Three-dimensional flow field variables ($p^*$, $u^*$ and $v^*$) are extracted at each sampling point. The 200 × 200 × 3 dimensional data that are extracted represent each instantaneous field. The obtained data is arranged in chronological order to obtain a dataset for training and testing the neural network. Each dataset including 1000 time-steps flow fields data.

After fixing the architecture of the neural network, we train the proposed hybrid deep neural networks in Tensorflow. As for how to choose parameters of the architecture of the neural network will be detailed in the next section. Learning rate values used in optimization are set as 0.0001 for fist 500 epochs, 0.00001 for second 500 epochs and 0.000001 for last 500 epochs. With the iteration of the losses and the weights in the network, this network gets closer to the real mapping relation. By back propagation to the weights in each layer, the loss function which stands for the RMSE drops rapidly.

To achieve the goal of predicting the flow fields at future occasions based on flow fields at previous occasions continuously, we recycle the output of the trained network to its input and updated the input recursively as the time-step advancement, with an initial condition taken from several snapshots of CFD data. So that, the network is able to achieve long-time predictions of flow fields even without known CFD data in the coming period.

*4.2. Parameters chosen of neural network architecture*

As explained in section *2.1*, the hybrid deep neural network is constituted by the convolutional neural network, improved convolutional Long-Short Term Memory neural network and deconvolutional neural network. It has been proven in previous work (Han et al., 2019) that six convolutional layers and six deconvolutional layers can accurately capture the spatial features of the flow fields and accurately reconstruct the high-dimensional flow fields. Therefore, in this article we focus on the influence of the LSTM layers' structure parameters on the prediction accuracy of the flow field around a forced oscillation cylinder. ConvLSTM layers' structure parameters mainly include the number of ConvLSTM cells per layer and the number of layers.

Figure 6 shows how the average RMSE in 150 time steps in the test phase varies with the number of ConvLSTM cells, when there is only one ConvLSTM layer in the deep neural network. It can be seen that the average root mean square error is the smallest when the number of cells is 3. It is best to predict flow field at next step by previous 3 time steps flow fields. This means the more time step input cannot make flow field prediction at the next moment more accurate. The flow field farther away from the moment to be predicted has less influence on the flow field at the moment to be predicted.

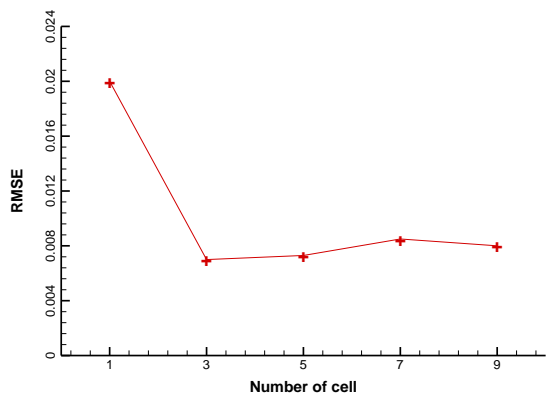

Figure 6. Comparison of average root mean square error under different cell numbers per layer

Then we discuss the impact of the number of ConvLSTM layers on the prediction accuracy by changing the number of ConvLSTM layers and setting the number of ConvLSTM cells per layer as 3. Figure 7 shows how the average root mean square error in 150 time steps varies with the number of ConvLSTM layers. It can be seen from the figure that the prediction error increases with the number of layers. Increasing the number of LSTM layers is to capture deeper information. But the structure of the flow field around a moving cylinder is relatively simple. And more layers increase the neural network variables, which may lead to greater errors. So that for this case one LSTM layer is enough to capture the spatial-temporal features of unsteady flows around a forced oscillation cylinder. For the problem with more complicated flow field characteristics, perhaps more ConvLSTM layers are needed to predict the flow field at future occasions more accurately.

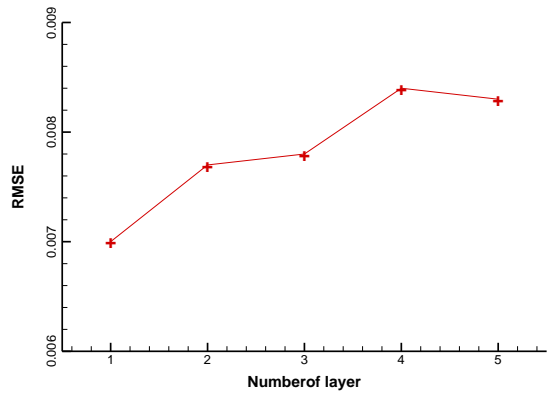

Figure 7. Comparison of average root mean square error under different ConvLSTM layers numbers

*4.3. Time series prediction at Re = 100*

In this section, the flow around moving cylinder cases with amplitudes A= 0.25D, 0.3D, 0.35D, 0.4D at Re = 100 are employed as the dataset for training the hybrid deep neural network to capture the underlying dynamics, while the flow around moving cylinder cases with amplitudes A= 0.425D is used for testing. The hybrid deep neural network is constituted by six CNN layers, one improved ConvLSTM layer with three cells and six DeCNN layers. Figure 8 shows that the training error defined by Eqn. (4) decrease with the increasing number of training steps. After 500000 training steps, the training error converges to less than 0.006. The average RMSE between the predicted and accurate results of the flow fields in 150 time steps in the test phase is less than one percent.

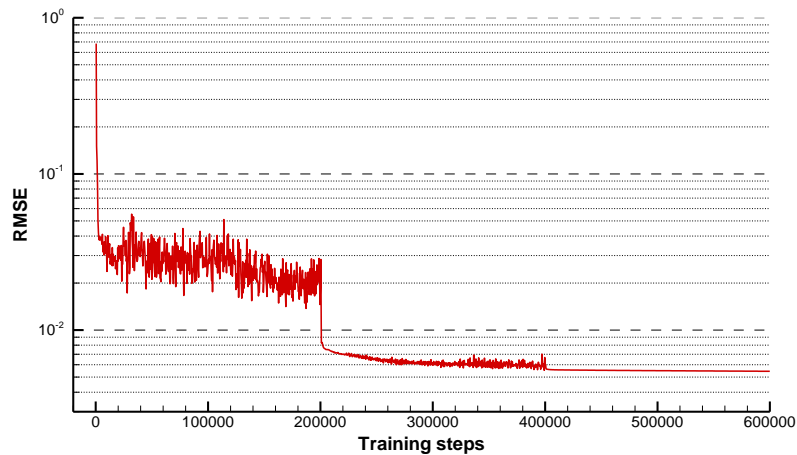

Figure 8. Training error decrease with the increasing number of training steps.

Comparisons of instantaneous flow fields between the hybrid deep network predicted results and CFD results in 20 time-steps are shown in Figure 9, Figure 10 and Figure 11. It should be clarified that all predictions after $3\delta t$ are predicted based on the hybrid deep neural network previous prediction without any CFD data. From comparisons, we can get that flow fields predicted are found to agree well with CFD simulation flow fields. One characteristic position, red star shown in Fig. 5, is selected to show the time series prediction accuracy. The spatial coordinates of the point are described by dimensionless x* and y*: circle center (0, 0); red star point in the wake A (3.25, 0.5). Time series of the three flow fields variables at selected positions, predicted by the network and calculated by CFD, are compared in Fig. 12. From comparisons, we can get that the flow fields predicted by the network still show good agreement with the CFD results in 150 time-steps. It prove that each part of the neural network structure has completed the predetermined target and the amplitude effects have been learned by the hybrid deep neural network. The results also show that the hybrid deep neural network has good extrapolatory capability and prediction error does not increase with time.

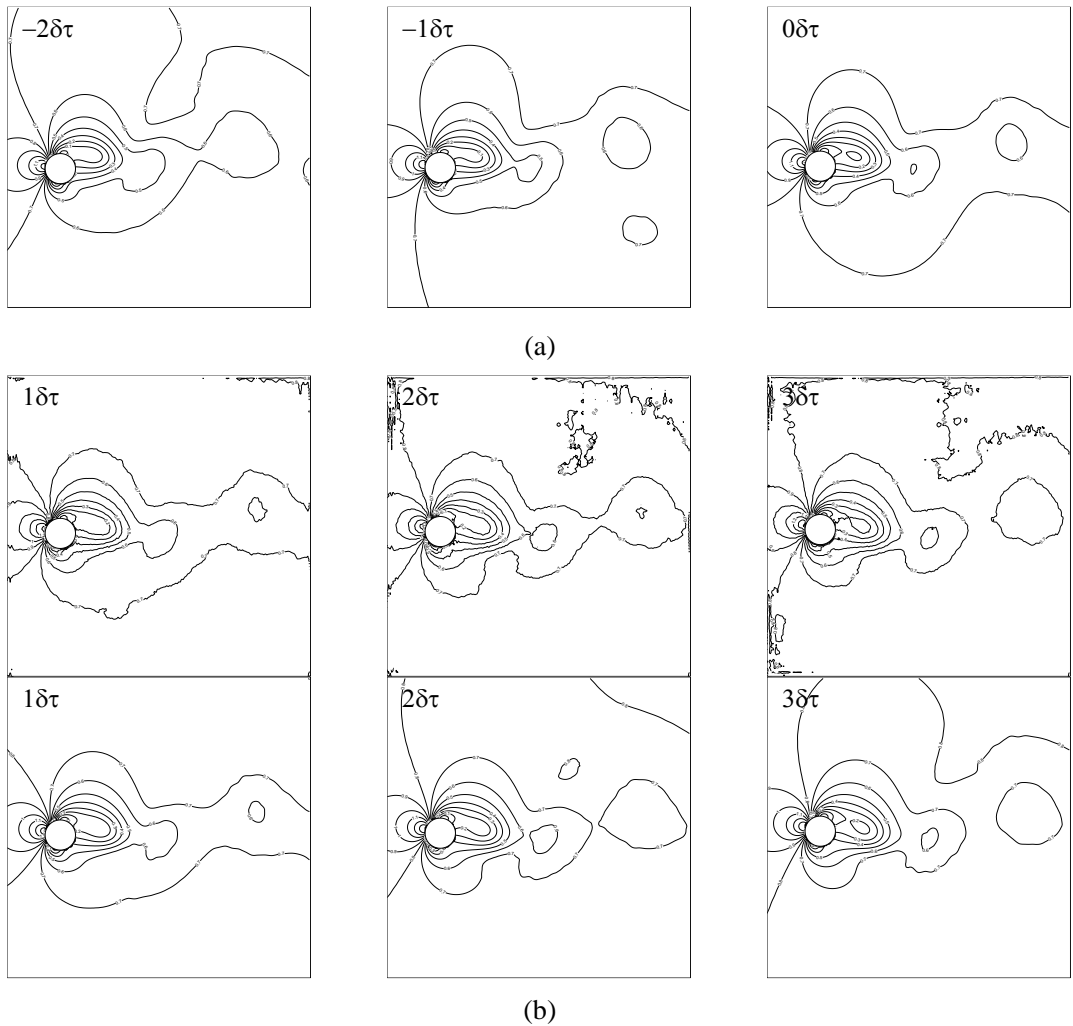

(a)

(b)

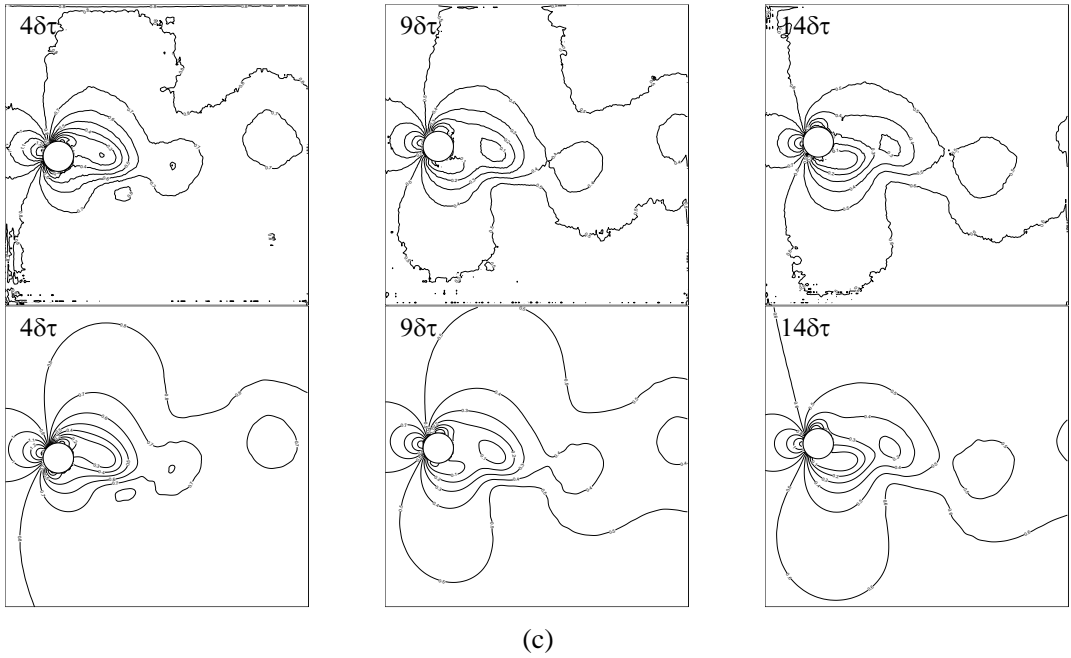

(c)

Figure 9. Comparisons of pressure instantaneous fields between model predictions and CFD results. (a) Input set; (b) Comparisons of every single step (1δt), the first row is the model predictions, the second row is the CFD results; and (c) Comparisons of every five step (5δt)

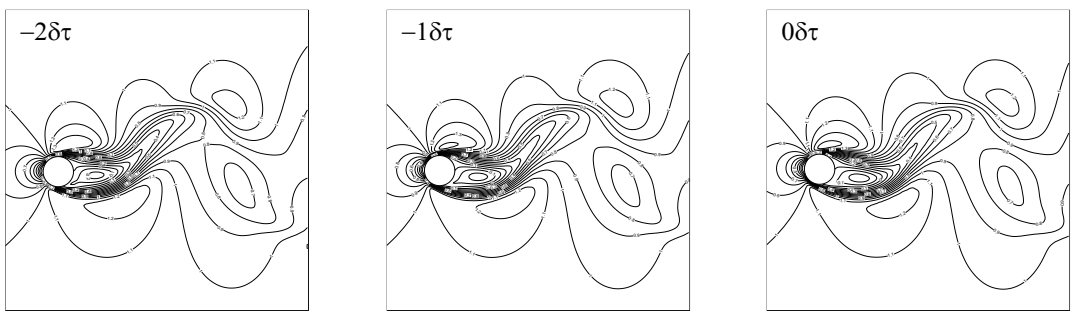

(a)

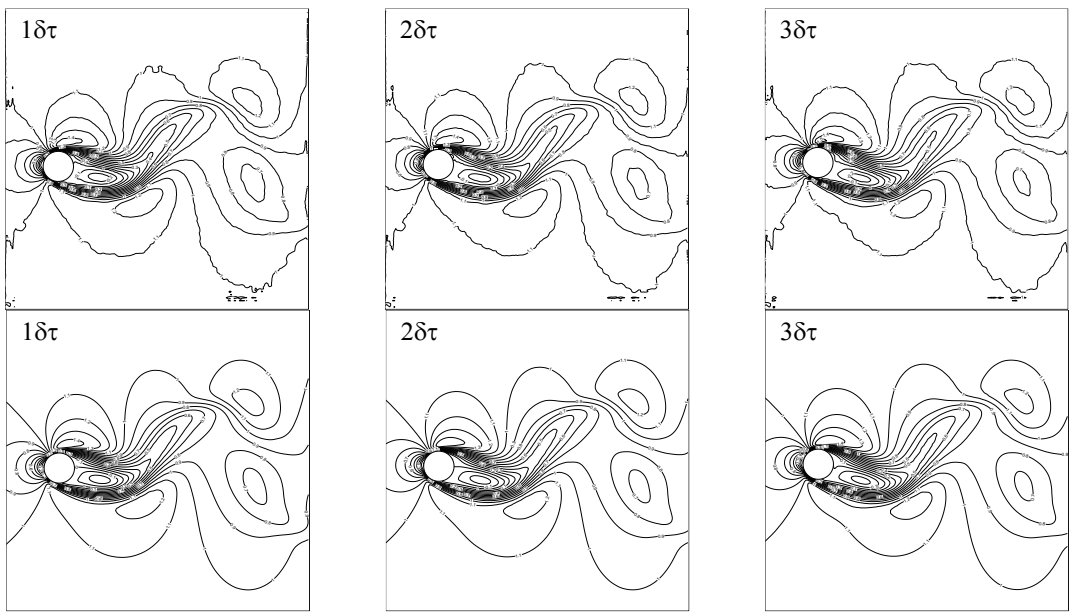

(b)

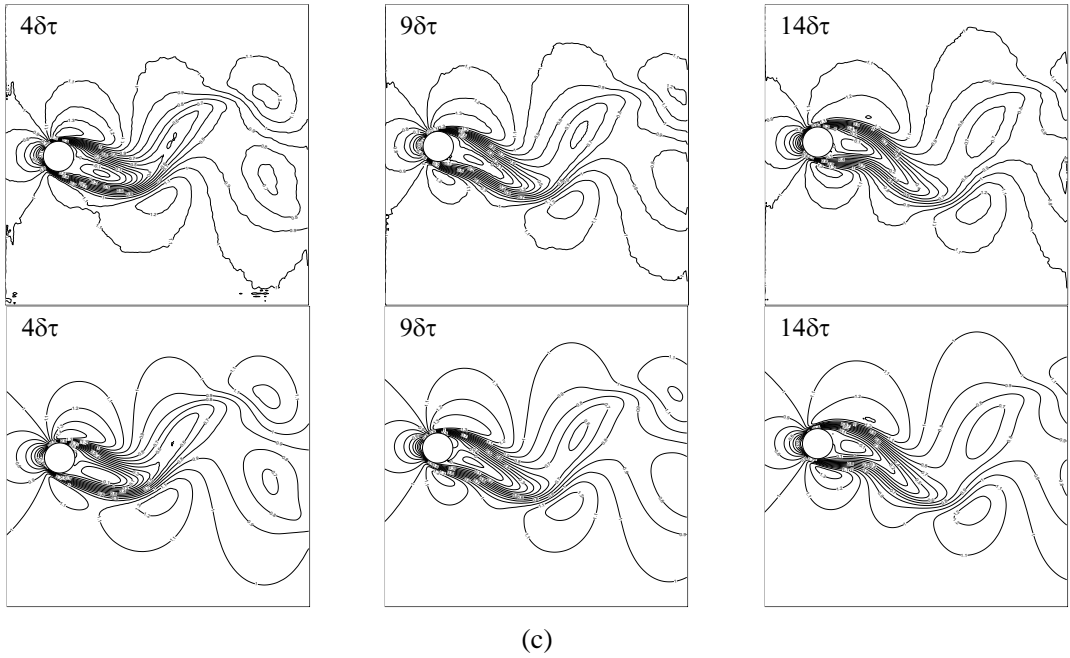

(c)

Figure 10. Comparisons of streamwise velocity instantaneous fields between model predictions and CFD results. (a) Input set; (b) Comparisons of every single step (1δt), the first row is the model predictions, the second row is the CFD results; and (c) Comparisons of every five step (5δt)

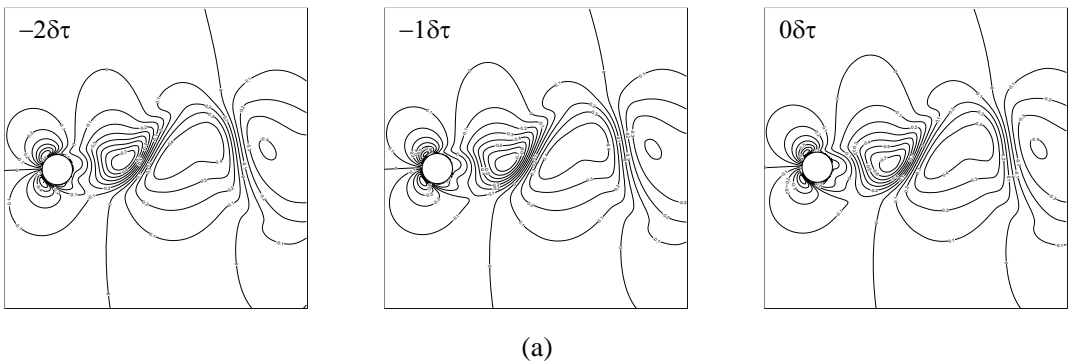

(a)

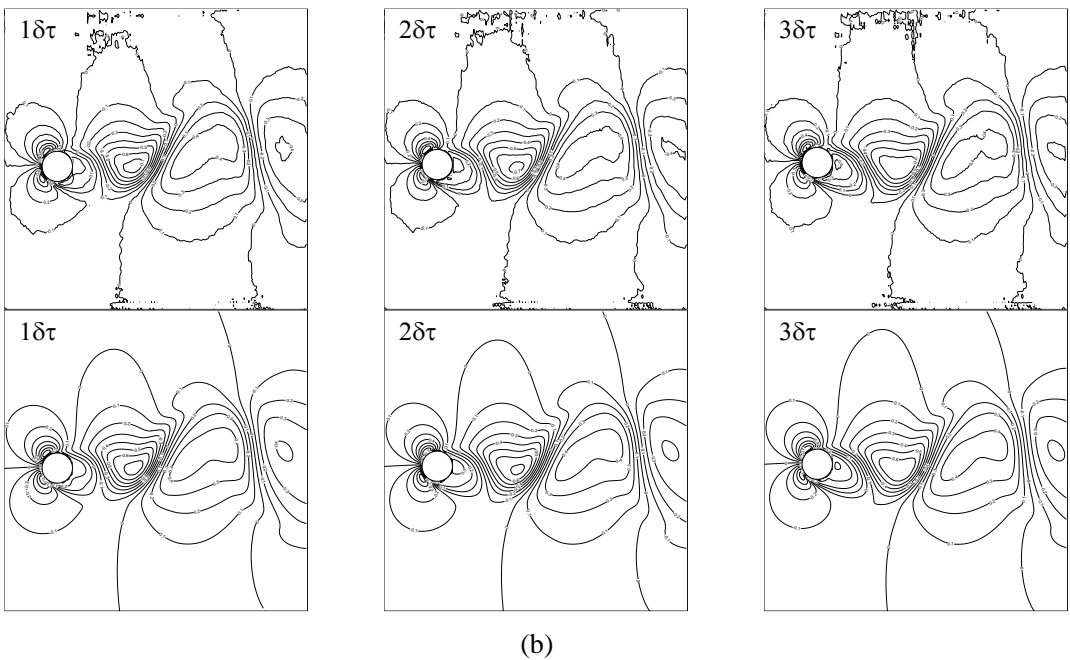

(b)

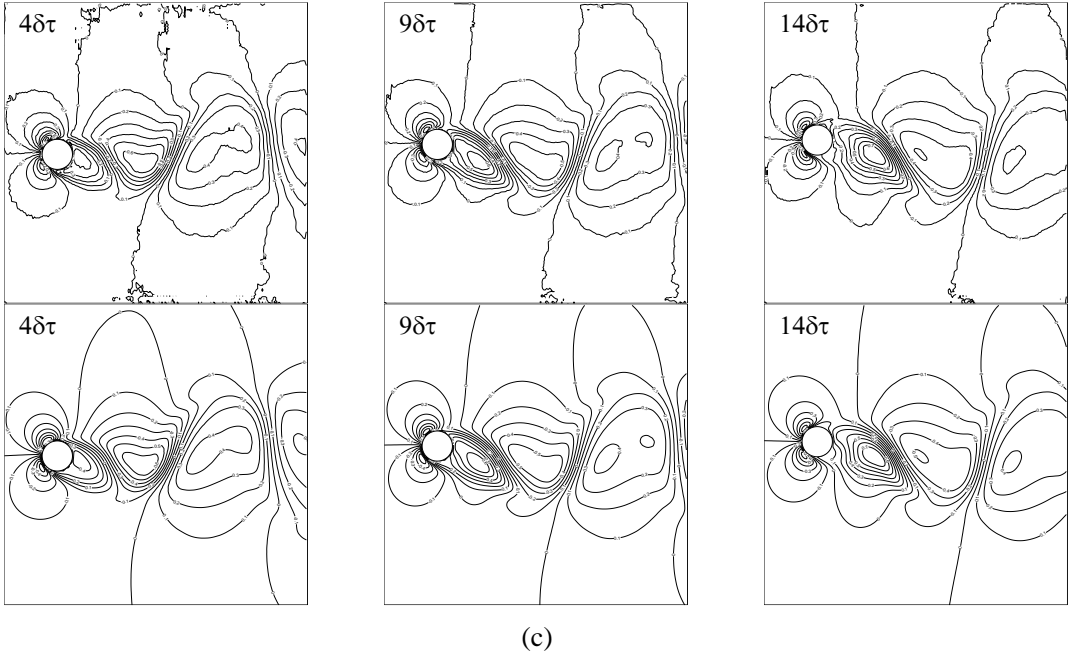

(c)

Figure 11. Comparisons of vertical velocity instantaneous fields between model predictions and CFD results. (a) Input set; (b) Comparisons of every single step (1δt), the first row is the model predictions, the second row is the CFD results; and (c) Comparisons of every five step (5δt)

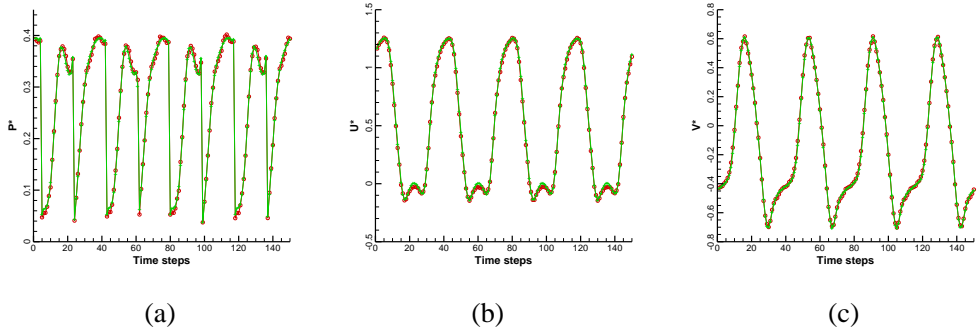

Figure 12. Comparisons of flow variables time histories in the wake between the model predictions and CFD results, (a) pressure, (b) streamwise velocity, (c) vertical velocity. Red circle shape point, model prediction at point A; green plus shape point, CFD results at point A.

### 4.4. Time series prediction at Re = 200

The deep neural network structure parameters are optimized when the Reynolds number is 100. In this part, we will test whether these parameters are effective for another situation. Same as the last experiment, the flow around moving cylinder cases with amplitudes A= 0.25D, 0.3D, 0.35D, 0.4D at Re = 200 are employed as the dataset for training the hybrid deep neural network to capture the underlying dynamics, while the flow around moving cylinder cases with amplitudes A= 0.425D is used for testing. The hybrid deep neural network is same too. The training error defined by Eqn. (4) decrease with the increasing number of training steps. After 500000 training steps, the training error converges to less than 0.01. The average RMSE between the predicted and accurate results of the flow fields in 150 time steps in the test phase is less than two percent.

Comparisons of instantaneous flow fields between the hybrid deep network predicted results and CFD results in 20 time-steps are shown in Figure 13, Figure 14 and Figure 15. Time series of the three flow fields variables at selected positions, predicted by the network and calculated by CFD, are compared in Fig. 16. From comparisons, we can get that the flow fields predicted by the network still show good agreement with the CFD results in 150 time-steps. It means that the amplitude effects have been learned by the hybrid deep neural network. Since the flow at Re = 200 is more complicated, the average RMSE of predictions at Re = 200 is not good enough as predictions at Re = 100. But the comparisons between the hybrid deep network predicted results and CFD results show that this kind hybrid deep neural network is able to capture the spatial-temporal features from high-dimensional unsteady flow fields at different Reynolds number. If you want more accurate prediction, you need to optimize structure parameters of the hybrid deep neural network.

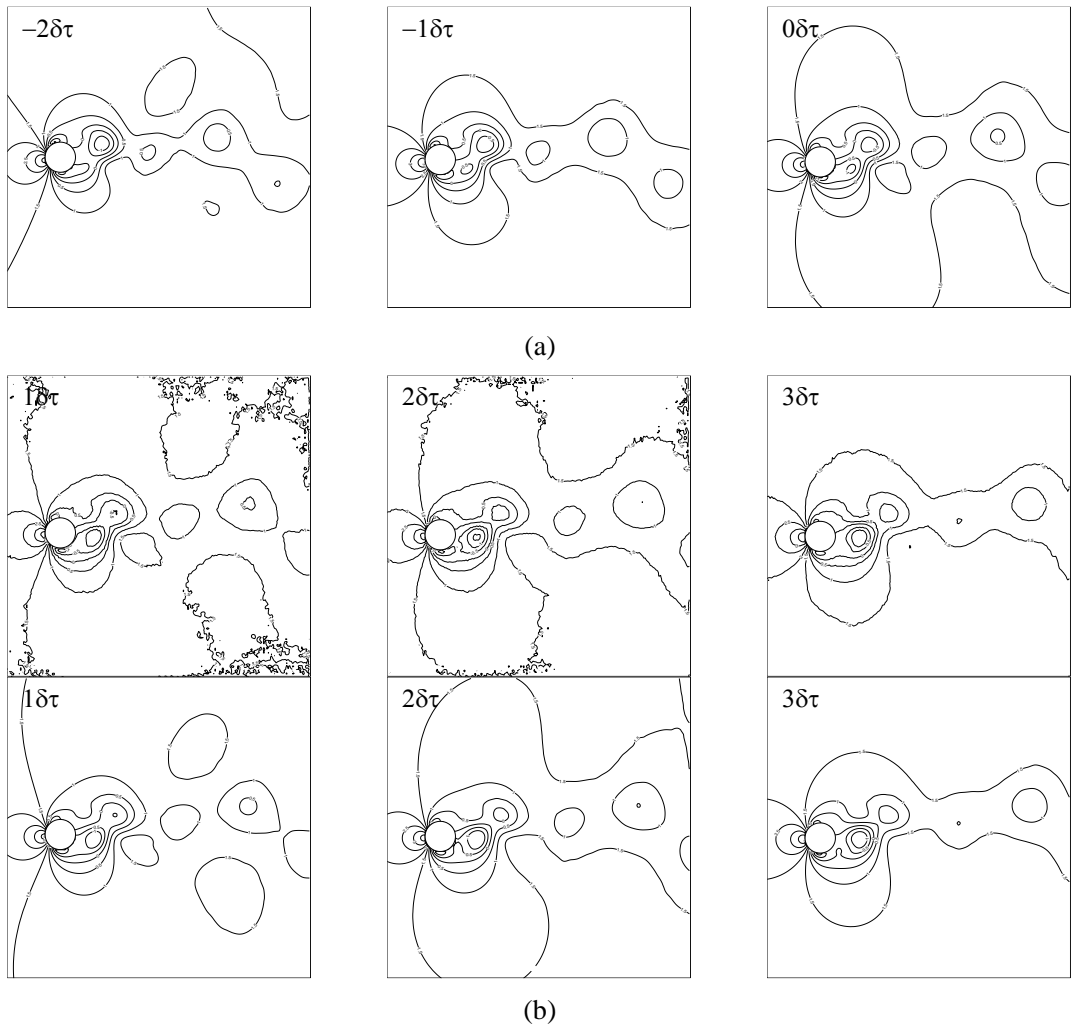

(a)

(b)

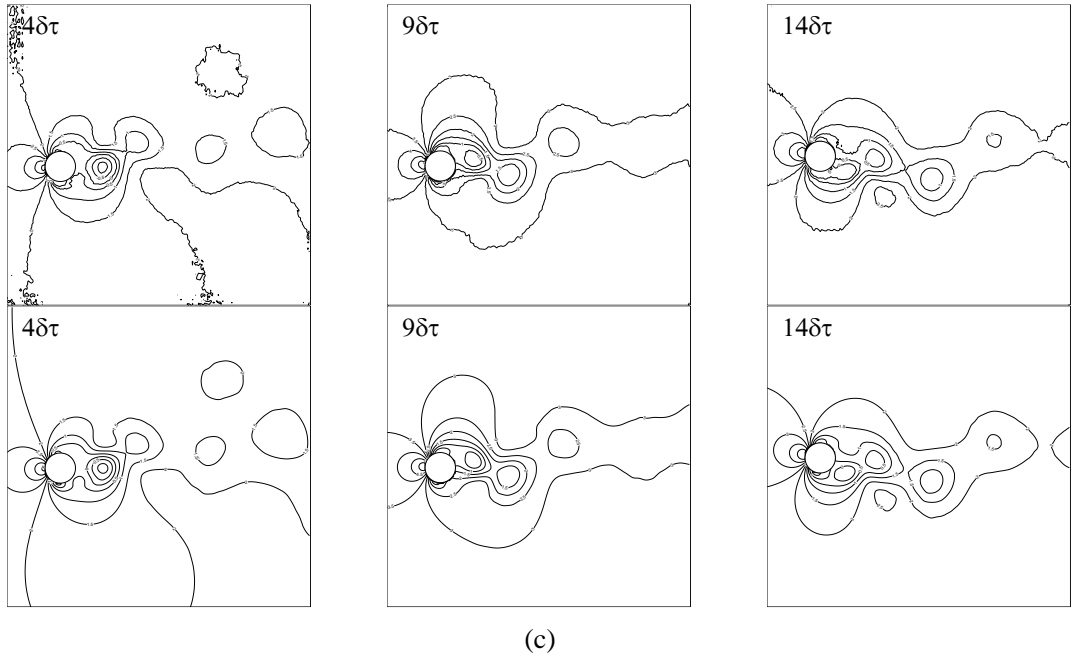

(c)

Figure 13. Comparisons of pressure instantaneous fields between model predictions and CFD results. (a) Input set; (b) Comparisons of every single step (1δt), the first row is the model predictions, the second row is the CFD results; and (c) Comparisons of every five step (5δt)

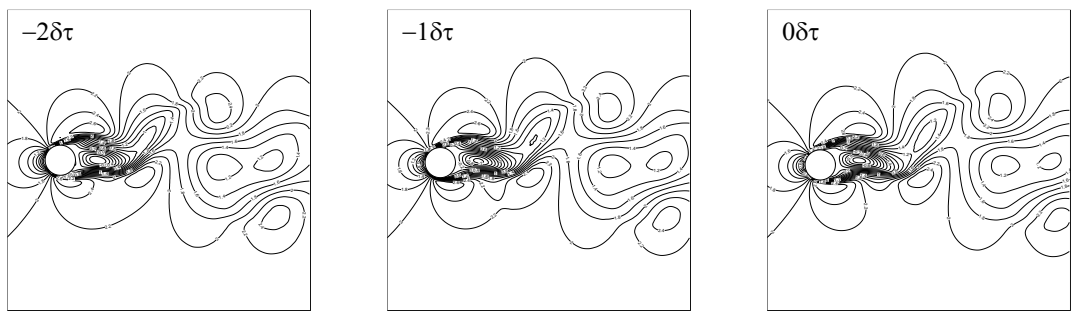

(a)

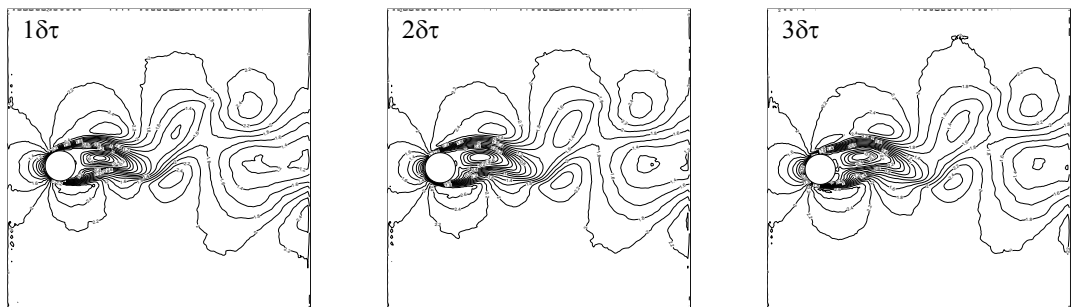

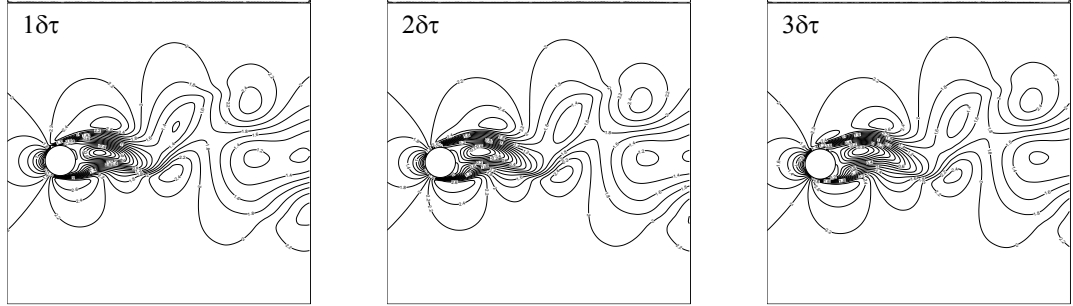

(b)

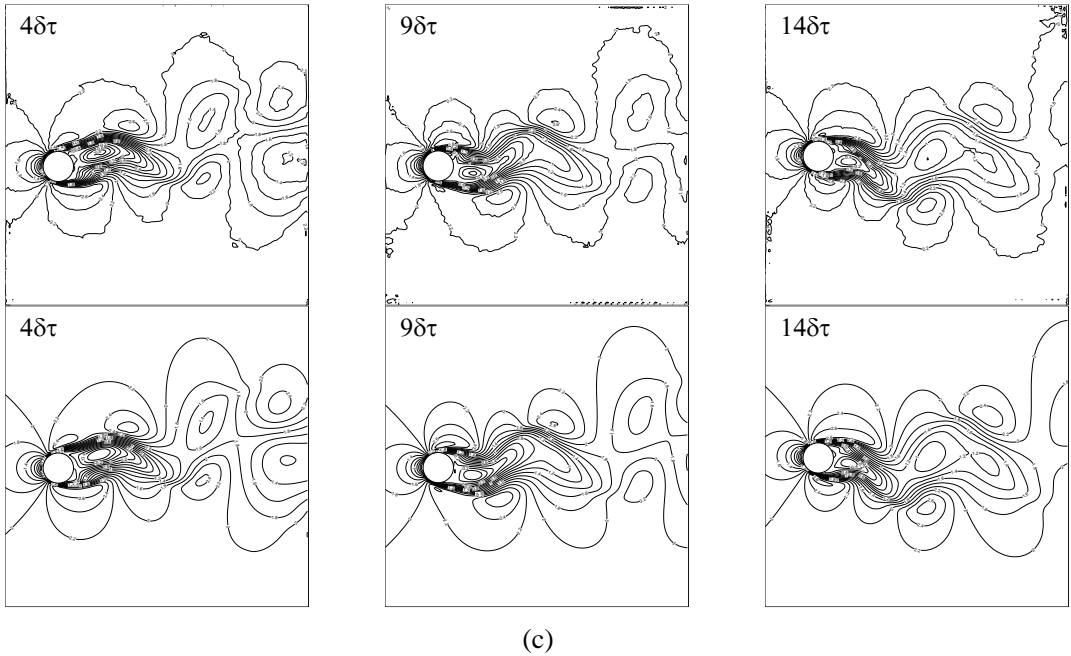

(c)

Figure 14. Comparisons of streamwise velocity instantaneous fields between model predictions and CFD results. (a) Input set; (b) Comparisons of every single step (1δt), the first row is the model predictions, the second row is the CFD results; and (c) Comparisons of every five step (5δt)

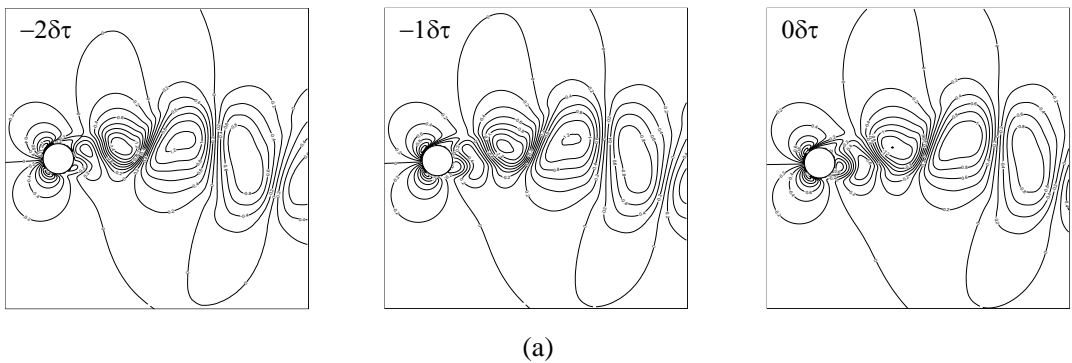

(a)

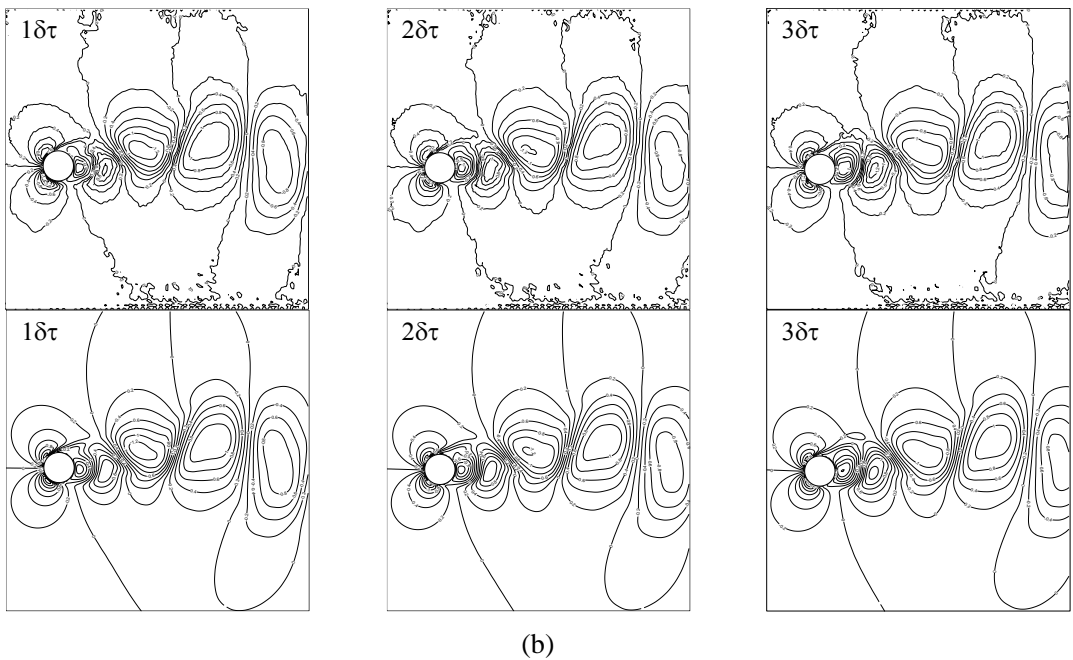

(b)

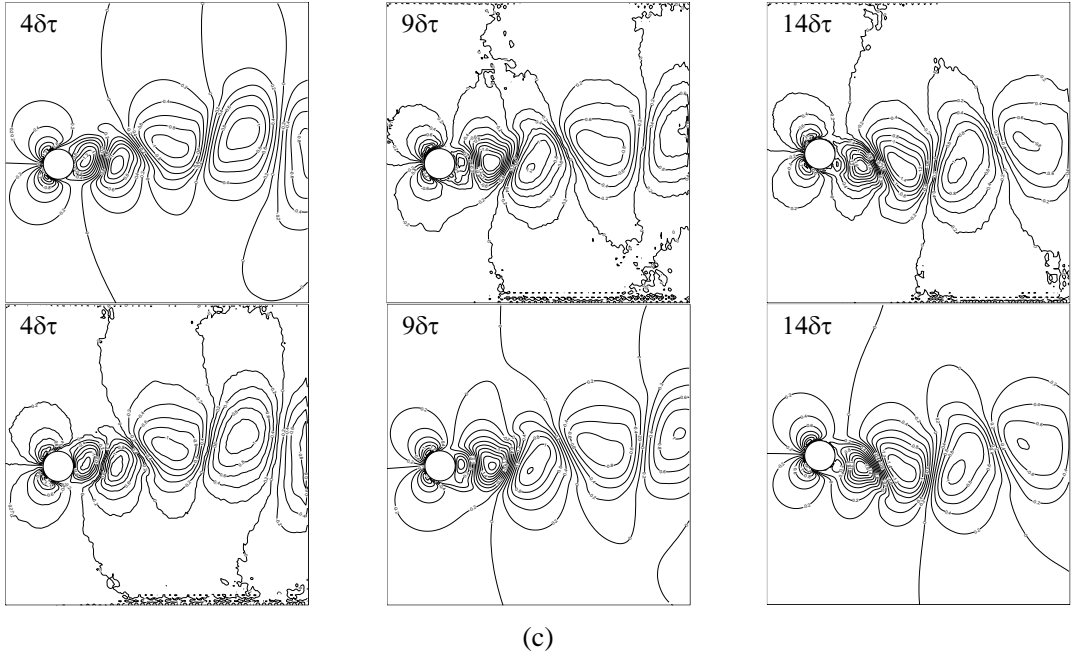

(c)

Figure 15. Comparisons of vertical velocity instantaneous fields between model predictions and CFD results. (a) Input set; (b) Comparisons of every single step (1δt), the first row is the model predictions, the second row is the CFD results; and (c) Comparisons of every five step (5δt)

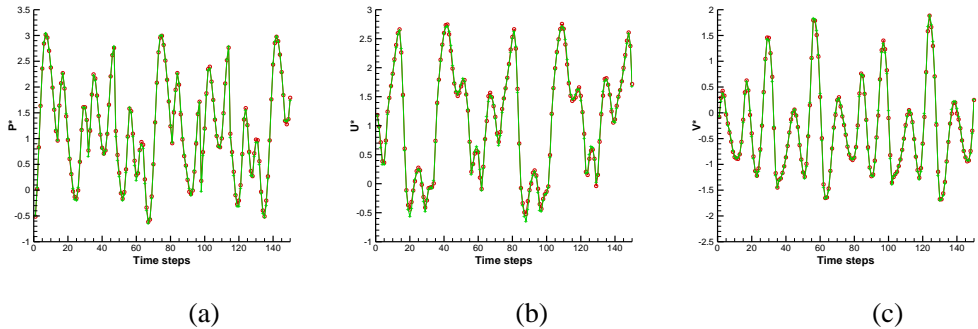

(a)          (b)          (c)

Figure 16. Comparisons of flow variables time histories in the wake between the model predictions and CFD results, (a) pressure, (b) streamwise velocity, (c) vertical velocity. Red circle shape point, model prediction at point A; green plus shape point, CFD results at point A.

## 5. Conclusions

A novel hybrid deep neural network architecture was designed to capture the spatial-temporal features directly from high-dimensional unsteady flow fields around moving boundaries. The hybrid deep neural network is constituted by CNN, improved ConvLSTM and DeCNN. The CNN layers are designed to capture the complex mapping directly from high-dimensional input flow fields and represent it in low-dimensional form. The ConvLSTM layer is designed to capture temporal features between low-dimensional features representation and boundary motion information then predict the low-dimensional features of flow fields at future occasions. In order to adapt to the problem of flow field prediction with moving boundaries, we modified ConvLSTM cell making it able to learn temporal features from flow fields and boundary position change. The DeCNN layers are used to represent the predicted low-dimensional features to high-dimensional output fields, with the same dimension as input fields. This kind of deep neural network can capture accurate spatial-temporal information from the spatial-temporal series of unsteady flows around moving boundaries.

The flow around a forced oscillation cylinder at various amplitudes are carried out to establish the datasets training the networks. The trained hybrid deep neural networks are then tested by the prediction of the flow field at future occasions whose amplitude is out of the range of testing dataset. The predicted flow fields using the trained hybrid deep neural networks are in good agreement with the flow fields calculated directly by computational fluid dynamic solver. The amplitude effects were learned by the hybrid deep neural network. This hybrid deep neural network can achieve fast and accurate prediction of unsteady flow fields with moving boundaries, which is very important to flow control and aerodynamic optimization application. The new prediction method can be used in flow control for vibrating cylinder, where the fast high-dimensional nonlinear unsteady flow calculation is needed.

The proposed hybrid deep neural network shows good potential in modeling spatial-temporal features of flow fields around moving boundaries, however there is still much work to be further done. Now it can only be used to solve the problem of prediction unsteady flows around forced vibration moving boundaries. The proposed new method is also expected to be used to deal with the problems of fluid-structure interactions and flow control for vibrating cylinder in future. In next step, we will focus on how to predict the surface pressure through the wake field information by deep neural network, so as to obtain the lift force and drag force. Finally realize the fast calculation of the end-to-end flow-solid interactions problem using the deep neural network completely.

## Acknowledgements

This work was partially supported by the National Natural Science Foundation of China (No.11872293,11672225), Science and Technology on Reliability and Environment Engineering Laboratory (No.6142004190307), and the Program of Introducing Talents of Discipline to Universities (No. B18040).